\documentclass[twocolumn]{aastex62}
\usepackage{natbib}
\usepackage{graphicx}	% Including figure files
\usepackage{amsmath}	% Advanced maths commands
\usepackage{amssymb}	% Extra maths symbols
\usepackage{cases}
\usepackage{array,multirow}
\usepackage{upgreek}
\usepackage{textcomp}
\usepackage{hyperref}
\usepackage[referable]{threeparttablex}
\usepackage{booktabs, dcolumn}
\usepackage[normalem]{ulem}
\usepackage{booktabs,tikz}

\graphicspath{{./}{../pictures/}}

\received{}
\revised{}
\accepted{}
\submitjournal{ApJ}

\shorttitle{Tidal disruption events}
\shortauthors{Krolik et al.}

\makeatletter
\newcommand*{\rom}[1]{\expandafter\@slowromancap\romannumeral #1@}
\makeatother

\newcommand{\beq}{\begin{equation}}
\newcommand{\eeq}{\end{equation}}
\newcommand{\simlt}{\mathrel{\hbox{\rlap{\hbox{\lower4pt\hbox{$\sim$}}}\hbox{$<$}}}}
\newcommand{\simgt}{\mathrel{\hbox{\rlap{\hbox{\lower4pt\hbox{$\sim$}}}\hbox{$>$}}}}

\newcommand{\physrad}{\mathcal{R}_{\rm t}}

\newcommand{\Lphysr}{\mathcal{L}_{\rm t}}

\def\apjl{ApJL}
\def\apj{ApJ}
\def\mnras{M.N.R.A.S.}
\def\aap{A\&A}
\def\nat{Nature}
\def\aplett{ApL}

\def\apjs{ApJ Supp.}

\def\prd{prd}

\definecolor{pinegreen}{rgb}{0.0, 0.4, 0.0}
\definecolor{vc}{rgb}{0.55, 0.,0.55}
\definecolor{as}{rgb}{1., 0.65,0.79}
\definecolor{violet}{rgb}{0.54, 0.17, 0.89}
\definecolor{cb}{rgb}{0., 0.18, 0.39}

\defcitealias{Ryu1+2019}{Paper 1}
\defcitealias{Ryu2+2019}{Paper 2}
\defcitealias{Ryu3+2019}{Paper 3}
\defcitealias{Ryu4+2019}{Paper 4}

\begin{document}

\title{Tidal Disruptions of Main Sequence Stars - V. The Varieties of Disruptions}
\correspondingauthor{Julian Krolik}
\email{jhk@jhu.edu}

\author{Julian Krolik}
\affiliation{Physics and Astronomy Department, Johns Hopkins University, Baltimore, MD 21218, USA}
\author{Tsvi Piran}
\affiliation{Racah Institute of Physics, Hebrew University, Jerusalem 91904, Israel}
\author[0000-0002-0786-7307]{Taeho Ryu}
\affil{Physics and Astronomy Department, Johns Hopkins University, Baltimore, MD 21218, USA}

\begin{abstract}
Tidal disruption events (TDEs) are generally imagined as the complete disruption of a star when it passes close to a supermassive black hole.  Relativistic apsidal precession is thought to quickly ``circularize" the bound debris, forming a compact accretion disk, which then emits a flare of standardized lightcurve and spectrum.  We show here that this picture holds in only a minority of cases.  TDEs are more diverse and can be grouped into several categories distinguished by stellar pericenter distance $r_p$; we  estimate the relative frequency of these categories.  Rapid circularization is rare both because it requires $r_p \leq R_{\rm circ} \sim  10r_g$ ($r_g \equiv GM_{\rm BH}/c^2$) and because most events with $r_p \leq 14r_g$ lead to  direct capture.  For larger pericenter distances, $R_{\rm circ} < r_p < 27r_g$ (for $M_{\rm BH}=10^6M_\odot$), main sequence stars with $M_* \lesssim 3$ are completely disrupted, but the bound debris orbits are highly eccentric and possess semimajor axes $\sim 100\times$ the scale of the expected compact disk.  Partial disruptions with fractional mass-loss $\gtrsim 10\%$ occur with a rate similar to that of total disruptions; for fractional mass-loss $\gtrsim 50\%$, the rate is $\approx 1/3$ as large.  Partial disruptions---which must precede total disruptions when the stars' angular momenta evolve in the ``empty loss-cone" regime---change the orbital energy by factors $\gtrsim O(1)$.  Partial disruption remnants are in general far from thermal equilibrium.  Depending on its orbital energy and conditions within the stellar cluster surrounding the SMBH, a remnant may return after $\sim O(100) - O(1000)$~yr and be fully disrupted, or it may rejoin the stellar cluster.

\end{abstract}


\section{Introduction} \label{sec:intro}

Tidal disruptions of main sequence stars by supermassive black holes were first discussed $\sim 40$~years ago \citep{Hills1976,FrankRees1976,LightmanShapiro1977,Lacy+1982}, and a characteristic shape for their lightcurves was proposed $\sim 30$~years ago: a sharp rise to a peak over roughly a month, followed by a decline $\propto t^{-5/3}$.  
This characteristic shape follows from the argument made by \citet{Rees1988} and \citet{Phinney1989} that the bound part of the debris stream returns to the pericenter at a rate $\propto t^{-5/3}$ supplemented by the assertion that upon return, the matter enters a compact accretion disk in which matter flows inward on a timescale faster than the mass-return timescale.  This disk is thought to form when relativistic apsidal precession leads to shocks capable of dissipating a large part of the matter's orbital kinetic energy near pericenter.

Several complications can make TDEs more diverse than this picture.  Events in which the pericenter distance $r_p \leq R_{\rm dc}$ lead to direct capture without tidal disruption; $R_{\rm dc} = 4r_g $ ($r_g \equiv GM/c^2$) for non-spinning black holes, and has only a very weak dependence on spin after averaging over orientation \citep{Kesden2012}.

At slightly larger radii, but only when $r_p$ is less than the radius $R_{\rm circ}$, apsidal precession can be strong enough to perform the role expected of it, driving rapid circularization and compact accretion disk formation. In Schwarzschild spacetime, the apsidal direction of highly-eccentric orbits is rotated by $\sim  (10r_g/r_p)$~radian per pericenter passage, suggesting that $R_{\rm circ} \sim 10r_g$; in Kerr, orbital plane precession may both diminish the probability of close-in debris stream intersection and delay stream intersections of all sorts \citep{Dai+2013,Guillochon.Ramirez-Ruiz2015}. However, black hole spin appears to have little effect on the energy distribution of the debris unless $r_p$ is very close to the direct capture distance \citep{Tejeda+2017,Gafton2019}.  See also the appendix for several other issues regarding black hole spin.

Because $\physrad$, the physical tidal radius within which stars are fully disrupted, can be considerably larger than $R_{\rm circ}$ ($\physrad \simeq 27 r_g$ for $M_* \lesssim 3$ when $M_{\rm BH} = 10^6M_\odot$; \citet{Ryu1+2019,Ryu2+2019}), 
for most TDEs in which the star is fully disrupted, quick formation of a compact accretion disk does not necessarily occur 
\citep{Shiokawa+2015,Guillochon.Ramirez-Ruiz2015}.  
The outcome of these events is still debated   \citep[e.g.][]{Guillochon.Ramirez-Ruiz2015,Piran+2015,Krolik+2016,Bonnerot.2017}; 
see \citet{Bonnerot2020} and \citet{Roth2020} for recent reviews.

At even larger pericenter distances, stars undergo partial disruptions, events in which only a fraction of the star is disrupted, and a significant, but perturbed, remnant remains. These take place when $\physrad < r_p < \widehat R_t$, where 
we define $\widehat R_t$ as the maximum pericenter within which at least a few percent of the star's mass is lost.  The ratio $\widehat R_t/\physrad$ varies by factors of a few with $M_*$ and $M_{\rm BH}$; it is $\simeq 2$ for $M_*=1$ and $M_{\rm BH} = 10^6$ \citep{Ryu3+2019}.  From here on, all masses will be given in units of the solar mass and all stellar radii in units of the solar radius.

Flares from partial TDEs should differ from those of both varieties of full TDE because the energy distribution of the bound debris is narrow and offset from $E=0$, leading to a mass fallback rate that peaks somewhat later than for full disruptions and declines much more rapidly  \citep{Guillochon+2013,Goicovic+2019,Miles+2020,Ryu3+2019}.   In addition, the remnant, which is initially much more extended and rotates much more rapidly than a main sequence star of its mass, can suffer quite disparate fates depending on how quickly it cools and the size of its orbit's semimajor axis \citep{Ryu3+2019}.

Finally, we note that if the SMBH is accreting before the star approaches, even at a slow rate, the interaction of the returning streams with the pre-existing accretion disk adds an entirely different aspect of diversity \citep{Chan2019,Chan2020}. In this case, the most dramatic effect of the encounter is a rapid infall of the inner accretion disk onto the SMBH.  The observed light curve then bears no resemblance to the classical $t^{-5/3}$ shape. 
%addtp - maybe we can add here See \cite{Bonnerot2020} for further complications. 

It is the thrust of this paper to expand upon these distinctions based on $r_p$ and to link (somewhat speculatively) the dynamical contrasts they imply with observational contrasts.  In order of their size, which is also the order in which we will discuss them, we call these different regimes: ``circularized total disruptions" ($R_{\rm dc} < r_p \leq R_{\rm circ}$); ``common total disruptions" ($R_{\rm circ} < r_p \leq \physrad$); ``partial disruptions" ($\physrad < r_p < \widehat R_t$); and ``unconventional total disruptions".
Note that ``circularization" in this context means not only that because of rapid energy dissipation the original eccentricity of the debris orbits is substantially reduced; it also means rapid formation of a compact disk whose inflow time is shorter than the mass-return timescale.
The last term refers to a total disruption of a partial disruption remnant in which the remnant did not have time to relax to a steady state configuration before returning to the black hole. Although the transition between the different regimes is not perfectly sharp, the key properties vary rapidly enough with $r_p$ to make these distinctions useful.

 In a number of respects, the categories we call ``circularized total disruptions" and ``common total disruptions" map onto what are called ``prompt" and ``slowed" events by \cite{Guillochon.Ramirez-Ruiz2015}.  As we develop our ideas, we will point out the degree to which our analysis agrees with, or extends, or disagrees with the arguments of this earlier paper.

The structure of the paper is as follows.  In \S \ref{sec:back} we discuss some of the relevant theoretical background and summarize some results from recent numerical simulations. 
We then turn to discuss different varieties of TDEs in order of increasing pericenter distance beginning with ``circularized total disruptions" in \S \ref{sec:circularized}, continuing with ``common total disruptions" in \S \ref{sec:common}, turning to partial disruptions in \S \ref{sec:partials} and concluding in \S  \ref{sec:un} with ``unconventional total disruptions". In each section we discuss first the basic dynamics, then speculate about their distinctive observational signatures, and lastly estimate the relative event rate of the regime in question.

\section{Background} 
\label{sec:back}

\subsection{Mass fallback rate and lightcurves}

The argument that the bolometric luminosity as a function of time is $\propto t^{-5/3}$ for times after the peak fallback rate begins with an estimate of the energy distribution in the debris resulting from the disruption:  spread symmetrically around zero with a half-width $\Delta\epsilon \sim GM_{\rm BH} R_*/r_t^2$, where $M_{\rm BH}$ is the mass of the black hole, $M_*$ and $R_*$ are the mass and radius of the unperturbed star, and $r_t \equiv R_*(M_{\rm BH}/M_*)^{1/3}$ is an order of magnitude estimate for the distance at which tidal forces are important.   The period of orbits with binding energy $\Delta \epsilon$ is $t_0 \simeq 0.11 M_*^{-1}R_*^{3/2} M_{\rm BH,6}^{1/2}$~yr. This estimate explains the month-timescales.  From here on, all masses will be given in units of $M_\odot$, all stellar radii in units of the Solar radius, and $M_{\rm BH,6} \equiv M_{\rm BH}/10^6$.

Given an angular momentum consistent with an orbit of pericenter $r_p \sim r_t$ and an orbital energy of the estimated scale, the orbits of the bound debris must be highly eccentric: $1 - e \leq 2 (M_*/M_{\rm BH})^{1/3} = 0.02 M_*^{1/3} M_{\rm BH,6}^{-1/3}$.  The rate at which mass returns to the star's pericenter immediately follows from a chain rule differentiation once the distribution of mass with energy, $dM/dE$, is determined.  \citet{Rees1988} argued that $dM/dE$ should be flat between $-\Delta \epsilon$ and $+\Delta\epsilon$, and detailed calculations employing realistic main-sequence internal density profiles for the stars have confirmed that this is a reasonable zeroth-order approximation \citep{Goicovic+2019,Golightly2+2019,LawSmith+2019,LawSmith+2020,Ryu2+2019}.

To predict the resulting lightcurve requires translating this mass-return rate to a dissipation rate coupled to a photon diffusion rate. 
This is often done supposing that relativistic apsidal precession would create strong shocks within the returning debris.  Having dissipated much of the orbital energy into heat, this material would then settle into a hot accretion disk with an outer radius $\sim 2r_p$.  The time for matter to accrete all the way to the black hole's innermost stable circular orbit (ISCO) would then be short compared to $t_0$, so the luminosity would follow the return rate if the accretion flow is radiatively efficient. In what follows, we will discuss the degree to which this view is upheld by subsequent work.

\subsection{Characteristic angular momenta}

Parabolic stellar orbits can be characterized by a single parameter because their pericenters are functions of angular momentum alone.  
In Schwarzschild spacetime, the relation between angular momentum and pericenter for a parabolic orbit is
\begin{equation}\label{eq:angmom}
L^2(r_p) = 2[(r_p/r_g)^2/(r_p/r_g - 2)] (r_g c)^2;
\end{equation}
spin corrections to this relation are relatively small, $\sim (a/M)/(L/r_g c)$.

In addition, as we discuss below, event rates are more directly related to
the stellar specific angular momentum $L_*$ than to the pericenter $r_p$.
It is therefore convenient to define the angular momenta of orbits with the pericenters separating the relevant TDE  categories: $L_{\rm dc} \equiv L(R_{\rm dc})$, $L_{\rm circ} \equiv L(R_{\rm circ})$, $\Lphysr \equiv L(\physrad)$, and $L_{\rm partial} \equiv L(\widehat R_t)$.

For orbits in the equatorial plane of a spinning black hole, $L_{\rm dc} = 2(1 + \sqrt{1 - a/M})r_g c$; averaging over all orbital orientations in Kerr spacetime, $\langle L_{\rm dc}\rangle \simeq 4r_g c$ with very small error \citep{Kesden2012}.
On the other hand, if $R_{\rm circ} \simeq 10r_g$ as estimated above, it is a reasonable approximation to adopt the Schwarzschild value for $L_{\rm circ} = 5r_g c$ because the fractional change due to spin is only $\sim 0.2 (a/M)$ (if $R_{\rm circ} > 10 r_g$, it is an even better approximation). In the following, we will use $R_{\rm circ} = 10r_g$ as a fiducial value;  
 in Sec.~\ref{subsec:structure} and the appendix we will discuss the degree to which it depends on parameters, including black hole spin.

As shown by \cite{Ryu1+2019,Ryu2+2019} for  middle-age main-sequence stars, $\physrad \approx [23.5(M_{\rm BH}/10^6)^{-2/3} + 3.5] r_g$, almost independent of $M_*$.
The corresponding $\Lphysr \approx 58 (r_g c)^2$ for $M_{\rm BH}=10^6 $ follows from Equation~\ref{eq:angmom}.   Stars of different ages have somewhat different internal structures, but  our  choice of middle-age main-sequence structures  corresponds reasonably well to the population mean.  Departures can be estimated from the work of 
\cite{LawSmith+2019,LawSmith+2020}, who considered different age main-sequence stars. Their central densities at different ages correspond, using the semi-analytic model for $\physrad$ of \cite{Ryu1+2019}, to a maximal change in $\physrad$ from zero-age main sequence to terminal-age main sequence (occurring for $M_*=1$) of a decrease in $\physrad$ by $\simeq 12\%$ and a 6\% decrease in $\Lphysr$.

\subsection{Rates}

The functional relation between event rate and event pericenter depends on how the surrounding stars evolve in angular momentum. Most calculations in the literature \citep[e.g.,][]{StoneMetzger2016} assume that the evolution is dominated by individual gravitational interactions with other stars, but collective mechanisms such as resonant relaxation \citep{RauchTremaine1996,RauchIngalls1998} or triaxiality in the stellar cluster \citep{Merrittbook2013} may also operate.  No matter which mechanism dominates, if the stars' orbital dynamics evolve in a fashion such that the change in angular momentum per stellar orbit is large compared to the angular momentum associated with $r_p = r_t$, their evolutionary regime is called ``full loss-cone" or ``pinhole" \citep{FrankRees1976,LightmanShapiro1977}.  In this regime, the stellar angular momentum distribution is smooth across the range of angular momenta associated with tidal disruption events.  The rate of an event with $r \leq r_p$ is therefore proportional to
the solid angle of the loss-cone in angular momentum space,
which is $\propto L^2(r_p)$.

The rate per black hole of TDEs of a given class in this orbital-evolution regime is
\begin{equation}\label{eq:rate_eqn}
S = \left[L_{\rm max}^2 - L_{\rm min}^2\right] \int \, dE_* \, \frac{\partial^2 f}{\partial L_*^2\partial E_*}(L_*=0,E_*) P_*^{-1}(E_*),
\end{equation}
where $L_{\rm max}$ and $L_{\rm min}$ define the range of angular momentum in which this class of event occurs, $df/dL_*^2$ is the distribution function of stars in the surrounding cluster with respect to the square of the stellar angular momentum, $E_*$ is the stellar specific energy, and $P_*(E_*)$ is the star's orbital period.

The situation is very different in the empty loss-cone regime, in which the rate of angular momentum evolution is so slow that every time a star passes through apocenter, its squared angular momentum changes by an amount $\Delta L_*^2 \ll |L(\physrad)|^2$. 
In this regime, stars undergo a random walk in angular momentum space until they find themselves with $L_*^2$ slightly less than $\Lphysr^2$, and are then destroyed by a total disruption.  In this limit, the rate of total disruptions is $\propto \ln \Lphysr$ \citep{LightmanShapiro1977}, augmented at about the 10\% level by occasional strong encounters \citep{Weissbein2017}.

\section{Circularized Total Disruptions}
\label{sec:circularized}
\subsection{Structure}
\label{subsec:structure}
At radii slightly larger than $R_{\rm dc}$, but only when $r_p$ is less than the radius $R_{\rm circ}$, apsidal precession can be strong enough to perform the role expected of it, driving rapid circularization and compact accretion disk formation. In Schwarzschild spacetime, the apsidal direction of highly-eccentric orbits is rotated by $\sim  (10r_g/r_p)$~radian per pericenter passage, suggesting that $R_{\rm circ} \simeq 10r_g$.   Here we show that this estimate is consistent with other considerations and discuss its uncertainty.

To ``circularize" in the sense of making the eccentricity small, the debris binding energy must increase by a factor $\sim (M_{\rm BH}/M_*)^{1/3} \sim 10^2 M_{\rm BH,6}^{1/3} M_*^{-1/3}$.  By analyzing shock dissipation when the impact is due to apsidal precession in Schwarzschild spacetime, \cite{Dai+2015} showed that even when $r_p$ is as small as $\approx 10 r_g$, the fraction of the orbital energy dissipated may be too little to put the debris on a circular orbit.
 When this happens, the semimajor axis $a_{\rm orb}$ of the new eccentric orbit is considerably greater than the radius at which the shock takes place \citep{Dai+2015,Guillochon.Ramirez-Ruiz2015}.    In fact, the analytic results of \cite{Dai+2015}  can be fit to within $\lesssim 10\%$ for all $M_*$ and $M_{\rm BH}$ such that $M_{\rm BH}/M_* \gtrsim 10^5$ by the expression $a_{\rm orb}/r_g \simeq 90(r_p/10r_g)^{2.5}$ for $r_p/r_g \lesssim 20$.   In other words, even for $r_p \simeq 10r_g$, the eccentricity after dissipation in a stream-intersection shock is still $\simeq 0.9$.   In the appendix we show that Kerr contributions to apsidal precession do little to alter these results.

In TDE contexts, ``circularization" also entails an inflow time $t_{\rm in}$ short compared to the mass fallback time $t_0$. For cases like these, it is difficult to predict $t_{\rm in}$ with much confidence because much about accretion in highly-eccentric disks remains poorly-understood.   Although the linear growth rate of the magnetorotational instability is roughly the same as in a circular disk \citep{EccMRI2018}, its nonlinear development remains unknown: the time to saturation has yet to be determined, and the relation between the magnetic stress and local pressure likewise remains an open question, particularly because all properties can be expected to vary periodically (at least in a statistical sense) around an elliptical orbit.

Nonetheless, applying the circular $\alpha$-model to disks such as these, and taking $\alpha (h/r)^2 \sim 10^{-2}$, \cite{Guillochon.Ramirez-Ruiz2015} used Monte Carlo parameter space sampling including several consequences of black hole spin to distinguish where $t_{\rm in}/t_0$ was larger or smaller than unity (``slowed" or ``prompt" in their terminology).  Expressing their criteria in terms of $\beta \equiv r_t/r_p$, they  found the critical $r_p$ separating these regimes is $\approx 10r_g$ for $M_{\rm BH} \sim 10^6$.

Thus, achieving a disk that is approximately circular requires pericenters significantly smaller than $10r_g$, likely inside $R_{\rm dc}$.   On the other hand, if $r_p \sim 10r_g$, the resulting disks are quite elliptical, but may be small enough to accrete quickly.   If one adopts our fit for $a_{\rm orb}$ as a function of $r_p/r_g$ and estimates the inflow time as a factor $Q$ times the orbital period for $a_{\rm orb}$, one finds another estimate for $R_{\rm circ}$:
\begin{equation}\label{eq:Rcirc}
R_{\rm circ} \sim 10 (Q/100)^{-0.27} \Xi^{-0.4} M_*^{0.085} M_{\rm BH,6}^{-0.13} r_g.
\end{equation}
Here $\Xi \sim 1$ \citep{Ryu1+2019,Ryu2+2019} is the debris energy distribution half-width in units of $\Delta\epsilon$.
This sort of estimate also points to $R_{\rm circ} \sim 10r_g$ with very little sensitivity to $M_*$ or $M_{\rm BH}$; instead, its dominant uncertainty is due to $Q$, which depends on unknown eccentric accretion dynamics.  In Sec.~\ref{sec:circrate} we discuss this uncertainty's impact on rate estimates.

\subsection{Radiation}

When $r_p \lesssim R_{\rm circ}\simeq 10r_{g}$, the shock speed is of order the free-fall speed near $r_p$, so that much of the orbital kinetic energy is dissipated into heat.  The result is an extremely hot gas, one with internal energy per particle $\langle \epsilon \rangle \simeq  25 (r_p/10r_{g})^{-1} (v_s/v_{\rm ff})^2$~MeV, for shock speed $v_s$ and free-fall speed $v_{\rm ff}$ at the location of the shock. Supported partly by the pressure due to this internal energy and partly by rotation (by definition, the specific angular momentum of the matter is comparable to what is needed for a circular orbit at this radius), the debris should form a geometrically thick disk.

 At such a high temperature, radiation would completely dominate the internal energy, so the immediately post-shock thermodynamic temperature is $\ll \langle\epsilon\rangle$,
\begin{align}
T &\sim \left(\rho \langle \epsilon \rangle/m_p a_{\rm th}\right)^{1/4} \\ \nonumber
    &\sim 2 \times 10^7 \left[\rho_{-4}{ (10r_{g}/r_p)}\right]^{1/4} (v_s/v_{\rm ff})^{1/2}\hbox{~K},
\end{align}
where we have scaled to a debris density $\sim 10^{-4}$~gm~cm$^{-3}$ on the supposition that $\rho \sim \rho_* R_*/a_0 \sim \rho_* (M_*/M_{\rm BH})^{2/3}$, for $a_0$ the scale of the debris orbits' semimajor axes and $a_{\rm th}$ the Stefan-Boltzmann constant for radiation energy density.  Thermal radiation can balance the rate at which debris orbital energy is dissipated, so one might expect a surface temperature determined by their equilibration:
\begin{align}
T_s \sim  2 \times 10^6 & M_*^{0.17} M_{\rm BH,6}^{-5/8}  (r_p/10r_{g})^{-3/4}\\ \nonumber
   &\times(v_s/v_{\rm ff})^{1/2}(t/t_0)^{-5/12}\hbox{~K}.
\end{align}
The exponent of $M_*$ comes from an approximation to the main-sequence mass-radius relation $R_* \propto M_*^{0.88}$ \citep{Ryu2+2019}.

The associated luminosity would be
\begin{align}
L\sim  9\times 10^{45} & M_{*}^{0.68} M_{\rm BH,6}^{-1/2} (r_p/10r_{g})^{-1}\\ \nonumber
    &\times(v_s/v_{\rm ff})^2 (t/t_0)^{-5/3}\hbox{~erg~s$^{-1}$}.
\end{align}

This is $\sim 60 L_E$ for $M_{\rm BH} \sim 10^6$, while $L/L_E \propto M_{\rm BH}^{-3/2}$.  Energy derived from accretion can in principle increase this by a factor of order unity, but by no more than that, because this already represents an energy per unit mass $0.1 (10r_{g}/r_p) c^2$.  As the estimated temperature indicates, it would emerge primarily in the soft X-ray band.

It is also possible that the accretion time is short compared to the cooling time.  This could be the case if the only internal heat transport mechanism is photon diffusion \citep{Begelman1979}.  Whether this is so depends on the internal dynamics.  If MHD turbulence driven by the magnetorotational instability achieves the amplitude found in ordinary super-Eddington disks, photon diffusion may be supplemented by photons trapped in magnetically-buoyant bubbles \citep{Blaes2011,Jiang2014}.  On the other hand, because the angular momentum in the debris, $< 5r_gc$ by definition, is only slightly greater than the maximum angular momentum permitting passage through the event horizon for weakly-bound matter ($4r_gc$), comparatively weak internal stresses are required for accretion, so the amplitude of MHD turbulence may be relatively low---which would also limit the additional dissipation associated with accretion.

Super-Eddington luminosity may drive an outflow carrying a sizable fraction of the net accretion rate \citep{Dai2018,Jiang2019}.  If so, it would be optically thick out to a radius $\gg r_p$; as a result, the emerging spectrum would be diminished in energy by a factor $\sim (r_p/r_{\rm photo})^{1/2}$ for photospheric radius $r_{\rm photo}$ \citep{Strubbe.Quataert.2009,Dai2018}. So far simulations have not been able to make a clear prediction of $r_{\rm photo}$ or its dependence on system parameters.

After a few $t_0$, nearly all the bound mass will have returned to the black hole and joined the compact disk.  From this point on, it is no longer a disk in inflow equilibrium, i.e., one whose mass accretion rate at every radius matches the feeding rate at its outer radius (modulo local fluctuations).  Rather, it is a disk that is intrinsically in a state of inflow {\it disequilibrium} because it continues to accrete onto the black hole even while it no longer receives new mass at the outside \citep{Shen.Matzner.2014}.  Recent formal models of such a disk employing general relativistic dynamics and permitting non-zero stress at the ISCO (as found in global MHD simulations: \citet{Noble.et.al.2010,SK16,Avara2016}) lead to a slow decline in the total disk dissipation rate $\propto t^{-n}$ with $0.5 \lesssim n \lesssim 1$ \citep{Mummery2018,Mummery2019}.

\subsection{Rate}\label{sec:circrate}

\begin{figure}
\centering
\includegraphics[width=0.48\textwidth]{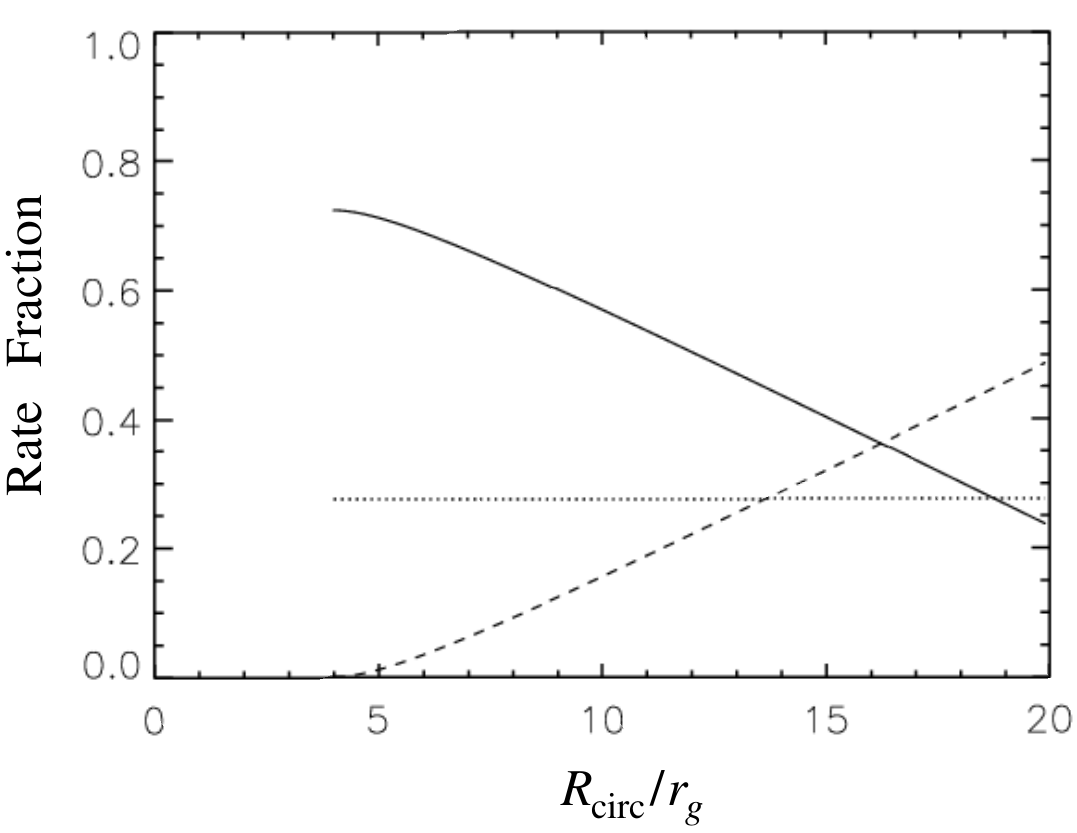}
\caption{The event rate fractions for total disruptions in the full loss-cone regime: common (solid), circularized (dashed), direct capture (dotted).  The normalization is for $M_{\rm BH} = 10^6$.}
\label{fig:eventrates}
\end{figure}

 Equation~\ref{eq:rate_eqn} gives the event rate when stellar dynamical evolution results in a full loss-cone.  To evaluate it, we need to identify $L_{\rm min}$ and $L_{\rm max}$.  In this context, $L_{\rm min}^2 = L_{\rm dc}^2 = (4r_g c)^2$, while $L_{\rm max}^2 = \min(L_{\rm circ}^2,\Lphysr^2)$. Consequently, $L_{\rm max}^2$ can never be more than $L_{\rm circ}^2$, which is $25 (r_gc)^2$ for $R_{\rm circ} = 10r_g$.  For this estimate of $R_{\rm circ}$, the fraction of all events with $r_p \leq R_{\rm circ}$ resulting in tidal disruptions rather than direct capture is always $\leq 0.36$, {\it wholly independent of both $M_*$ and $M_{\rm BH}$}.  For different choices of $R_{\rm circ}$, the ratio between circularized events and direct captures can be read off the curves shown in Figure~\ref{fig:eventrates}.  As this figure shows, circularized events are less likely than direct captures so long as $R_{\rm circ} < 14r_g$; they become extremely unlikely if $R_{\rm circ}$ is smaller than our fiducial estimate of $10r_g$. For $M_{\rm BH} = 10^6$, circularized events happen less often than common events for $R_{\rm circ} < 16r_g$. None of these values of regime-dividing pericenters depends significantly on black hole spin because, as remarked in Sec.~\ref{sec:back}, the spin contribution to the $L$--$r_p$ relation is $\sim (a/M)/(L/r_gc)$.   Whether $R_{\rm circ}$ is larger than $10r_g$ depends almost entirely on whether eccentric accretion disks can produce inflow at rates noticeably faster than circular disks (see eqn.~\ref{eq:Rcirc}).  The only significant dependence on stellar or black hole mass in these event fractions is that, with increasing $M_{\rm BH}$, $\Lphysr$ decreases, diminishing the maximum value of $R_{\rm circ}$ for which common events are the majority of total disruption flares.

The  situation for empty loss-cone evolution is different.  The rate of rapidly-circularizing total disruptions is an even smaller fraction of all total disruptions so long as $\physrad$ is significantly greater than $R_{\rm circ}$.  The only avenue to a rapid circularization event is through the $\sim 10\%$\footnote{Even for the 10\% that are strongly deflected, the relative rates of capture, circularized and common are still approximately those given in Figure~ \ref{fig:eventrates}.} probability channel provided by the tail of the stellar scattering angular momentum exchange distribution \citep{Weissbein2017}.

However, $\physrad \sim R_{\rm circ}$ when $M_{\rm BH} \gtrsim 5 \times 10^6$, so that the change in angular momentum required to take a star from its last value of $L$ for a partial disruption to $\leq L_{\rm circ}$ might be $ \ll \Lphysr$. For these higher-mass black holes, provided $-\Delta L_* < L_* - L_{\rm dc}$, empty loss-cone evolution leads almost exclusively to circularized total disruptions, modulo an exception to be discussed in \S~\ref{sec:un}. 
Of course, when $L_* < L_{\rm dc}$, even slow angular momentum evolution ultimately leads to capture rather than total disruption.

\section{Common Total Disruptions}
\label{sec:common}
\subsection{Structure}

In this class of events, the star is totally disrupted, but the stellar pericenter is too large for there to be strong apsidal precession.   As a result,  orbital intersection between debris streams with different return times occur near their orbital apocenters rather than near their pericenters \citep{Shiokawa+2015,Dai+2015,Guillochon.Ramirez-Ruiz2015}.   Because the eccentricity is so large, this means the kinetic energy available for dissipation is much smaller than if the intersection took place near pericenter, typically by a factor $\sim 100$.  Dissipating this amount of energy leaves most of the matter on orbits with semimajor axes $\sim 100r_p$, far outside $r_p$.

The subsequent evolution of such a system was explored in the simulation of \cite{Shiokawa+2015}.   Late-arriving streams still suffer an intersection shock, but a second shock also appears in the apocenter region.  Over a time $\simeq 10t_0$, these shocks gradually diminish in magnitude as the mass-return rate diminishes.
At this point, the debris mass resides in an asymmetric, highly elliptical flow that accretes onto the SMBH rather slowly; this is the regime called ``slowed" by \cite{Guillochon.Ramirez-Ruiz2015}.

When the shocks become weak, the only mechanism capable of transferring angular momentum is MHD turbulence stirred by the magnetorotational instability.  Its linear growth rate in these circumstances is comparable to that found in circular orbits \citep{Chan2019}; if its growth to nonlinear saturation takes $\sim 10$~orbits, as is the case for circular orbit flows, it does not begin until $\sim 10t_0$ after the disruption, and even when it does, the accretion time from these larger radii is likely to be a multiple of the orbital time, i.e., $ \gg t_0$.  Accretion of the majority of the bound mass may therefore be a relatively slow process. 

On the other hand, a non-negligible minority is deflected by the ``nozzle shock" that forms where the streams converge toward the orbital plane as they approach $r \sim r_p$.  This portion of the debris actually can enter a compact accretion disk.  However, this does not necessarily happen on the gas's first passage through pericenter; the entire process may require up to $\simeq 10t_0$ to go to completion \citep{Shiokawa+2015}.

\subsection{Radiation}

Such a dynamical situation has several implications regarding observable properties.  The apocenter-region shocks have total heating rates quite similar to the observed optical/UV luminosity ($\sim 10^{44}$~erg~s$^{-1}$) over the period $\simeq 3 - 10t_0$ after the star's pericenter passage; the warmed area is also similar  \citep{Piran+2015,Krolik+2016,Ryu+2020} to the size inferred from the temperature ($\sim 2-4 \times 10^4$~K) of the thermal spectrum seen in that band (e.g., the large sample reported in \citet{vanVelzen+2020}).  The orbital speeds in that region are likewise similar to the observed widths of emission lines.

Accretion through the compact disk produces a similar luminosity for about the same time, and with an effective temperature \citep{Krolik+2016} not far from the $\simeq 50 - 100$~eV commonly seen \citep{vanVelzen+2020}.  When the slowly accreting matter coming in from the apocenter region begins to arrive, one might expect a slow decay in the disk luminosity as well as a slow decrease in its characteristic temperature.

\subsection{Rate}

Once again, the rate of full loss-cone events is given by Equation~\ref{eq:rate_eqn}, but $L_{\rm max}^2 = \Lphysr^2$ and $L_{\rm min}^2 = L_{\rm circ}^2$.  For $M_{\rm BH} = 10^6$, the nearly $M_*$-independent value of $\physrad$ is $\simeq 27r_g$ \citep{Ryu1+2019,Ryu2+2019}; this translates to $\Lphysr^2 = 58 (r_gc)^2$, so that $\Lphysr^2 - L^{2}_{\rm circ} =33 (r_g c)^2$ for $R_{\rm circ} = 10r_g$.
Recalling the rate we estimated for circularized total disruptions in the previous section, we see that, for $M_{\rm BH} = 10^6$  and for $R_{\rm circ}=10r_g$, the rate of common total disruptions is, for full loss-cone evolution, $\gtrsim 4\times$ the rate of rapidly-circularizing total tidal disruptions. Reference to Figure~\ref{fig:eventrates} shows how this ratio changes as a function of $R_{\rm circ}$.

This result is in qualitative agreement with the work of \cite{Guillochon.Ramirez-Ruiz2015}, who  find that for  $M_{\rm BH} \lesssim 10^7$, the majority of TDEs are of this nature (see their fig. 7).  Unfortunately, we cannot compare more quantitatively because \cite{Guillochon.Ramirez-Ruiz2015} do not state the maximum stellar angular momentum they considered.

The rate of common total disruptions due to empty loss-cone evolution is the classical rate $\propto \ln \Lphysr$---but with a complication. To reach the edge of the loss-cone by this means, stars have no choice but to pass through the range of $L_*^2$ in which partial disruptions take place.  Consequently, they {\it must} lose mass as their angular momentum wanders through this region of phase space, and the fraction of the star's mass that is lost increases as its angular momentum nears $\Lphysr$.  As we will discuss at greater length in \S~\ref{sec:partials}, total disruptions reached in this way predominantly involve low-mass stars because even stars beginning the process with high-mass become low-mass by the time they are completely disrupted.  Further complications regarding orbital energy evolution and the thermal state of partial disruption remnants will also be discussed in $\S$~\ref{sec:partials}.

\section{Partial Disruptions}\label{sec:partials}
\subsection{Structure and radiation}

Like full disruption events, partial disruptions also produce a significant amount of debris, roughly half of which is bound and half unbound, even if the total mass in debris is, by definition, less than in a complete disruption.  However, unlike complete disruptions, the energy distribution of the debris is generically bimodal, rather than approximately flat \citep{Guillochon+2013,Goicovic+2019,Ryu3+2019}.  To the degree that less than all the star's mass is lost, there is always a depression in $dM/dE$ for $E \approx 0$.  This depression is deeper and wider for events with less mass-loss.  The immediate consequence is that the mass-return rate for partial disruptions declines more steeply after reaching its peak: as the fractional mass-loss diminishes from unity, the logarithmic slope of the post-peak fallback rate becomes significantly steeper than the -5/3 of complete disruptions { \citep{Guillochon+2013,Goicovic+2019,Golightly2+2019,Ryu3+2019,Miles+2020}}. For the same fractional mass-loss, the slope is shallower for low-mass stars than for high-mass stars, i.e., $M_* \gtrsim 1$ \citep{Ryu3+2019}. When $\lesssim 10\%$ of the star's mass is lost,  the decline steepens further.
In addition, the energy at which $dM/dE$ peaks is also somewhat smaller in absolute magnitude than the characteristic energy spread for full disruptions \citep{Ryu3+2019}.  It therefore takes longer for the debris stream to return to the SMBH vicinity, so that the peak mass return rate is reached somewhat later.  The magnitude of the mass return rate is also depressed, both because the return is slower and because there is less debris mass than in a full disruption.

Partial disruptions also differ from total disruptions in that they rapidly circularize {\it only} if $M_{\rm BH} \gtrsim 1 \times 10^7$.   Thus, their phenomenology should resemble that of common total disruptions, except for the adjustments to their mass return rate described in the previous paragraph.

\subsection{Rate}

As shown by \citet{Ryu1+2019,Ryu4+2019}, the relation between the fractional remnant mass left after a partial disruption of a main sequence star and the character of the star's original orbit is a nearly universal function, independent of $M_*$ and $M_{\rm BH}$, when phrased in terms of the angular momentum of the orbit.  To be specific, if we define the variable
\begin{equation}
x_L \equiv \frac{L_*^2 - L_{\rm dc}^2}{\Lphysr^2 - L_{\rm dc}^2},
\end{equation}
then
\begin{equation}\label{eq:mremfrac}
\frac{M_{\rm rem}}{M_*} = 1 - x_L^{-3}.
\end{equation}
The remnant left behind is within a few percent of the mass of the original star when $x_L > 3$.

\begin{figure}
\centering
\includegraphics[width=0.48\textwidth]{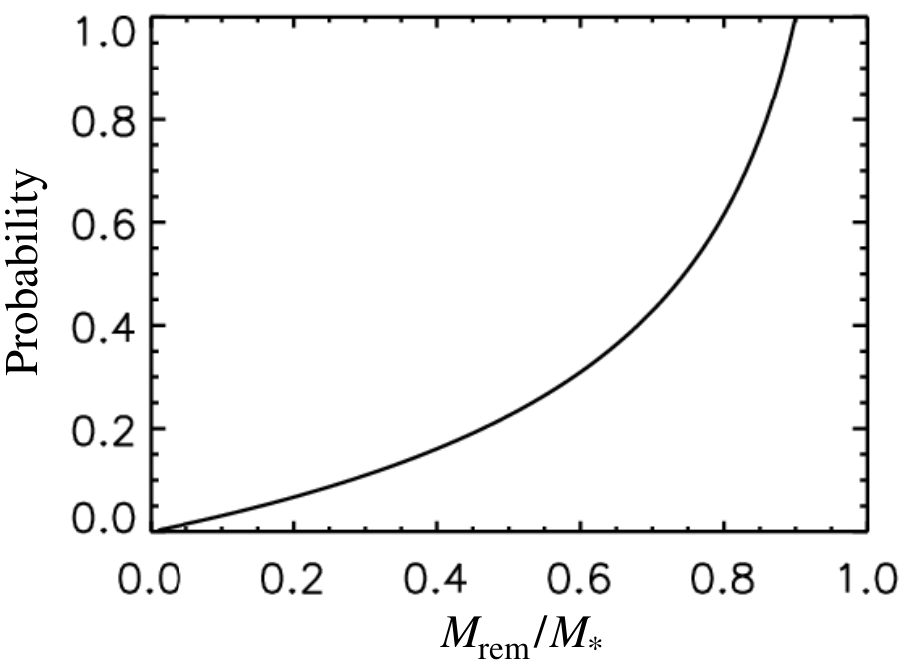}
\caption{The probability that a partial disruption event yields remnant mass fraction $\leq M_{\rm rem}/M_*$ as predicted by the relationship between orbital angular momentum and $M_{\rm rem}/M_*$ developed in Paper~1.}
\label{fig:partial_rate}
\end{figure}

These relations can be rearranged into the form
\begin{equation}
L_{\rm rem}^2 = L_{\rm dc}^2 + \frac{\Lphysr^2 - L_{\rm dc}^2}{\left(1 - M_{\rm rem}/M_*\right)^{1/3}}.
\end{equation}
Once again using the full loss-cone formalism of Equation~\ref{eq:rate_eqn}, the rate of events in which the remnant mass fraction is non-zero but $\leq M_{\rm rem}/M_*$ is then proportional to $L_{\rm rem}^2 - \Lphysr^2$.  This rate rises very steeply as a function of $M_{\rm rem}/M_*$.  For $M_{\rm BH} = 10^6$, $(L_{\rm rem}^2 - \Lphysr^2)/(r_g c)^2 \simeq 1.5$ for $M_{\rm rem}/M_*=0.1$, but rises to $\simeq 11$ for $M_{\rm rem}/M_*=0.5$, $\simeq 25$ for $M_{\rm rem}/M_*=0.75$, and $\simeq 50$ for $M_{\rm rem}/M_*=0.9$ (see Fig.~\ref{fig:partial_rate}).

If stars evolve in the full loss-cone regime and $M_{\rm BH} = 10^6$, the rate of events with fractional mass-loss $\geq 20\%$ is about the same as the rate of common total disruptions; the rate for events with a mass-loss fraction $\geq 10\%$ is $\simeq 1.5\times$ larger.  Because empty loss-cone evolution begins to depress the distribution function even for $L_*> \Lphysr$, the rate of events as a function of fractional mass-loss in this regime could have an even steeper dependence on $M_{\rm rem}/M_*$ than the curve shown in Figure~\ref{fig:partial_rate}. The fact that the probability for an event with a certain fractional mass-loss rises so sharply with decreasing mass-loss means that stars suffer many minor partial disruptions for every major one.  The most probable mass-loss history for a star evolving in this regime is therefore to lose most of its mass in a large number of weak events.   By the time such a star reaches the threshhold for a total disruption, its mass will, in most instances, be a fairly small fraction of its original mass. 

Partial disruptions do not only affect stellar mass: they also change the remnant's specific orbital energy.  In this way, they can also be viewed as accelerating energy evolution, and by that means, possibly even moving stars from one angular momentum evolution regime to the other.  In the sample of events examined by \citet{Ryu3+2019}, the change in energy due to the disruption is comparable to or larger than the typical kinetic energy of bulge stars.  When the star's incoming orbit is one that places it in the full loss-cone regime, its change in angular momentum per orbit $|\Delta L|$, although large compared to the tiny loss-cone angular momentum, could nonetheless still be small compared to the circular-orbit angular momentum $L_0$ corresponding to its energy.  Stellar interactions during a single orbit would then change its energy by a fractional amount comparable to $\max(|L_*|,|\Delta L|) |\Delta L|/L_0^2$.  By contrast, the partial disruption itself changes its energy by a factor of order unity or even greater, significantly accelerating energy evolution.  The conventional assumption of tidal disruption rate calculations that energy changes relatively slowly (e.g., as in \cite{StoneMetzger2016}) is therefore upended.  Because the contrast between empty and full loss-cone evolution is largely a matter of energy---higher energy orbits go out farther into the galaxy and have longer orbital periods, both promoting more interaction with other stars---stars can, as a result of a partial disruption, move from empty loss-cone evolution to full loss-cone evolution or vice versa in a single orbit.

Rapid energy evolution is especially important to stars on orbits placing them in the empty loss-cone regime because they {\it must} suffer partial disruptions before being totally disrupted.   Calculations of evolution in this regime must therefore account for rapid energy evolution: it is unavoidable.

\section{Unconventional total disruptions}
\label{sec:un}
\subsection{Structure and radiation}

In our discussion of the results of partial disruptions up to this point, we have implicitly restricted our analysis to the case in which the stellar remnant is able to relax thermally in less than a single orbital period, so that its structure has reverted to main-sequence by the time of its next pericenter passage.   That is not necessarily the case.  In the examples studied by \citet{Ryu3+2019}, the range of orbital periods for bound stars was from $\sim 400 - 40,000$~yr; for the shorter period stars,  even though the excess heat inherited by the remnant is concentrated in its outer layers, there might well not be time to relax thermally before returning to the black hole.  It is important to note in this regard that \citet{Ryu3+2019} found short orbital period remnants across the whole range of $r_p$ values yielding partial disruptions.

If a star does encounter the black hole for a second time while in a heated state, it will do so with essentially the same pericenter as before because it is exactly those stars with the shorter orbital periods that are least likely to suffer significant angular momentum change while traveling around an orbit.  When it does, it will have a core/halo structure: its core density may be several to tens of times smaller that that of the same-mass main sequence star, but it has an extended envelope whose density falls to much lower values and stretches out to several times the radius of the matched-mass main sequence star.  In addition, it can be spun up to a few tens of percent of break-up, especially in the case of severe mass-loss.

The semi-analytic model presented in \citet{Ryu1+2019} may provide some rough guidance to the fate of such a star.  This model, which reproduces the results of simulations reasonably accurately, predicts that the distance within which a black hole is capable of pulling off matter of a given density is $\propto \rho^{-1/3}$.  Thus, the primary criterion for a full disruption upon return is that the central density be reduced by a large enough factor (as compared to the original star) that the original pericenter, which produced only a partial disruption, is now within the physical tidal radius of the remnant.  Applying this criterion to the remnant sample of \citet{Ryu3+2019}, it appears that when a star has lost $\gtrsim 50\%$ of its mass, it will suffer a total disruption upon return if it returns before it relaxes to a more compact configuration.  It is plausible that this criterion might be loosened somewhat when, in addition, the remnant's rapid prograde rotation is taken into consideration; unfortunately, although there exist calculations of the debris energy distribution when such a star is totally disrupted \cite{Golightly2019,Kagaya2019}, the degree to which the physical tidal radius is altered has yet to be determined.  For those remnants with $M_{\rm rem}/M_* \gtrsim 0.5$, another partial disruption is more likely, but the mass-loss resulting it from should be considerably greater than if a main-sequence star of that mass passed at the same distance from the black hole.

The resulting flare from either a total or a partial disruption of such a distended remnant might have noticeably different properties from events due to tides exerted on a main sequence star.  The core/envelope configuration of an uncooled remnant is likely to lead to a rather different energy distribution for the debris mass, perhaps with both a greater concentration toward $E \approx 0$ and a larger spread in energy.  The energy spread could be further enhanced by prograde rotation \citep{Golightly2019,Kagaya2019}.  As a result, for both full and partial disruptions, the mass-return rate would exhibit an earlier rise and a more gradual decline at late times.  Moreover, because the pericenter distances at which these occur are in the range producing exclusively partial disruptions of main sequence stars, their debris can circularize rapidly only if the black hole is extremely massive.

\subsection{Rate}

The rate of unconventional total disruptions depends on how the remnant's orbital period ($\sim 10^2 - 10^5$~yr as found by \citet{Ryu3+2019}) compares to two other timescales: the remnant's thermal relaxation time and the time for dynamical interactions to alter the remnant's angular momentum by more than $\sim \Lphysr$.  If the orbital period is shorter than both, the remnant returns to the vicinity of the SMBH with essentially the same pericenter as before, but in a distended state. Preliminary estimates based on the semi-analytic model of \citet{Ryu1+2019} show that for $M_{\rm rem}/M_* \lesssim 1/2$, the remnant is sufficiently distended to suffer an unconventional total disruption.  Thus, greater mass-loss and shorter orbital periods (the latter for two reasons) favor unconventional disruptions at the next pericenter passage.

Clearly, the  rate of these events is most sensitive to the orbital period distribution of remnants and how, if at all, it is correlated with $M_{\rm rem}/M_*$.  The rate can be bracketed from above by the rate of partial disruptions (there is at most one such event per partial disruption). For the shorter orbital periods, the lower bound on the rate may be not much smaller.   For the longer orbital period portion of the distribution, the lower bound
could be much smaller. A detailed quantitative estimate of the rate will require detailed calculations of both the remnant thermal relaxation time and of the distribution of specific orbital energy with which remnants are ejected after a partial disruption.

\section{Discussion}

We have distinguished several different regimes of tidal disruption events according to their outcome.  The parameter with the greatest influence is the orbital pericenter or, equivalently and more physically informative, the star's orbital angular momentum.  The angular momentum has such central interest because it is the quantity most closely related to the rates of the different events.

\begin{figure}
\centering
\includegraphics[width=0.46\textwidth]{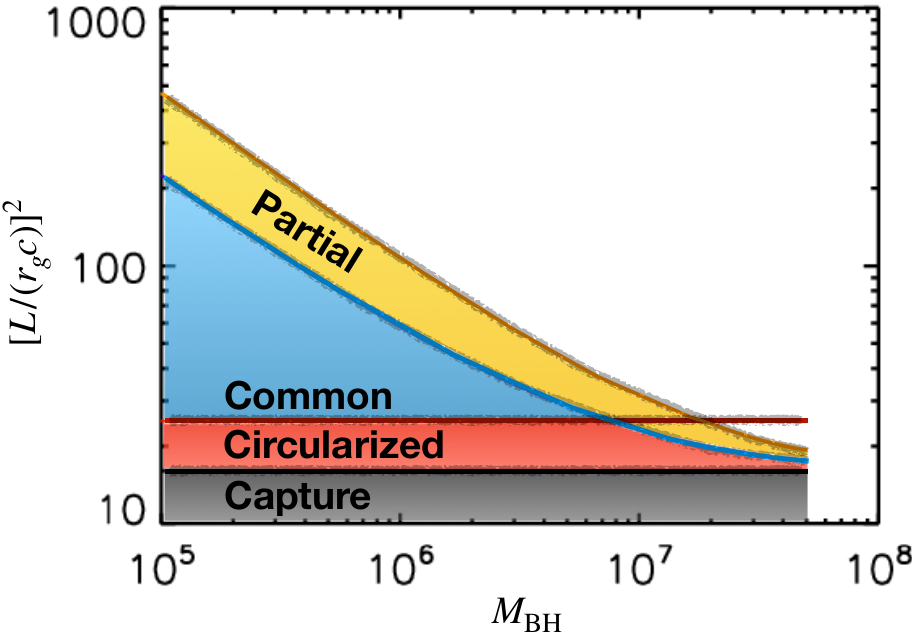}
\caption{Four critical values of $[L/(r_g c)]^2$: ($L_{\rm dc}/r_g c)^2\approx 16$ (black line); $(L_{\rm circ}/r_g c)^2 \approx 25$ (red line), taking our fiducial value for $R_{\rm circ}$; $(\Lphysr/r_gc)^2$, which is $\approx 60$ for $M_{\rm BH} = 10^6$  (blue curve), and $(L_{\rm partial}/r_g c)^2$ defined for $M_{\rm rem}/M_*=0.9$ (orange curve).Direct captures occur for angular momentum in the gray-shaded area below  the black line. Circularized total disruptions are found in the red-shaded area. Common disruptions occupy the blue-shaded area.  Partial disruptions with $\geq 10\%$ mass-loss are found in the yellow-shaded area. { The blurred boundaries between the regimes reflect the weak dependence on other parameters of the system, such as the stellar mass or the spin of the SMBH.} }
\label{fig:regimes}
\end{figure}

In Figure~\ref{fig:regimes}, partial disruptions with $\geq 10\%$ mass-loss occupy the yellow area; common total disruptions take place in the blue zone; circularized total disruptions happen in the red band; direct captures occur where the color is gray. If $R_{\rm circ}$ were different from $10r_g$, the red line would move up or down { in a way illustrated in Figure~\ref{fig:eventrates}}.

As { Figure~\ref{fig:regimes}} shows clearly, for small black hole mass, only a very small fraction of phase space is devoted to circularized total disruptions, and this fraction is made especially small by the fact that $(L_{\rm dc}/L_{\rm circ})^2$ is quite sizable for any $R_{\rm circ} \sim 10r_g$; it is 0.64 for our fiducial value.  For $M_{\rm BH} = 10^5$, the circularized fraction of all events within the physical tidal radius is only $\simeq 4\%$, and at $M_{\rm BH} = 10^6$, it is still only 15\%.

At higher black hole mass, 
$M_{\rm BH} \gtrsim 5 \times 10^6$, all total disruptions producing bright flares are circularized disruptions, but they represent $\leq 36\%$ of all events with $r_p \leq 10r_g$. As the black hole mass rises further, this fraction becomes even smaller as circularized disruptions become even rarer relative to direct capture.

\cite{Guillochon.Ramirez-Ruiz2015} also found that what they called ``prompt" events would be a minority of all TDEs for small black hole mass and dominate for large, but they overestimated the rate of these events by ignoring their suppression by direct capture.

 A corollary of the fact that circularized disruptions are a minority of all total disruptions { when} ${{M_{\rm BH} \lesssim 5 \times 10^6}}$ is that non-circularized events are the majority.   We have been able, for the first time, to make a clean estimate of how large this majority is because we have made use of the results of \cite{Ryu1+2019,Ryu2+2019,Ryu4+2019}.  Using general relativistic simulations of main-sequence stars with realistic internal density profiles for multiple values of $M_*$, they showed that the physical tidal radius, and therefore $\Lphysr^2$, is nearly independent of $M_*$ for $M_* \lesssim 3$.  Note that because their profiles were chosen for middle-aged stars, they effectively averaged over main-sequence ages.

For most of the relevant black hole mass range, partial disruptions (with mass-loss at least 10\%) should take place at a rate similar to that of common total disruptions. Their distribution in fractional mass-loss follows a universal form (eqn.~\ref{eq:mremfrac}).

These different varieties of tidal disruption should be identifiable through their differing observational properties.  Common total disruptions should produce thermal optical/UV emission at a temperature $\sim 2-4 \times 10^4$~K  \citep{Ryu+2020}, possibly accompanied by thermal X-rays with $kT \sim 50-100$~eV.  Their optical/UV lightcurves should show a rise to an approximate plateau, but decline roughly $\propto t^{-5/3}$ after $t \approx 10t_0$.   At still later times, the luminosity will depend on how rapidly matter can accrete from the debris apocenter region at $\sim 10^3 r_g$ to the ISCO region.  It is possible that the lightcurve at times $\gg t_0$ could be relatively shallow, with most of the light not emerging until long after the initial flare peak.  This last prediction may explain how it is that in a number of cases, both the UV and X-ray energy radiated 5--10~yr after TDE discovery was about the same as that seen in the first few months \citep{vanVelzen.2019,Jonker2019}.

Partial disruptions should resemble common total disruptions in terms of the initially radiated spectrum because the characteristic energy of the debris is, to within factors of 2, about the same.  However, because the mass-return rate falls off  more steeply, the initial plateau is likely to be shorter, and the decline following its end steeper, than in common total disruptions.  Like the common total disruptions, partial disruptions may also lead to an extended period of slowly-declining low luminosity emission.

Circularized total disruptions are comparatively rare and could be different. Their intrinsic emission should come at somewhat higher X-ray energies, and these events can produce substantial optical/UV luminosity only by reprocessing.  In addition, however, their significantly super-Eddington accretion rates may create qualitatively new phenomena,
perhaps explaining the small number of jetted TDEs \citep{Burrows11,Bloom+2011,Cenko+12}. 

The high mass accretion rate in a circularized disruption event (or in a TDE occurring where there is a pre-existing disk: \citet{Chan2019}) can form  a geometrically thick accretion disk.
\citet{CoughlinBegelman2014} suggested that such a disk could drive a relativistic jet by radiation pressure alone. 
Alternatively, given enough time, MHD turbulence could build up the magnetic field in the disk until its 
energy density per unit mass approaches the virial energy of the fluid.
Such a field can support a jet whose efficiency in Poynting radiation is $\sim O(1)$ in terms of accreted rest-mass when the black hole spins rapidly.  \citet{McKinney.et.al.2012}, for example, reported that in a disk whose initial field was toroidal, as would likely be the case for disks created by tidal disruptions, there were transient episodes of strong jets.  These might be enhanced by the fact that the orbital axes of TDE disks are in general oblique to the spin axis of the black hole, causing field toroidal with respect to the disk to have a poloidal component with respect to the black hole spin.

Unconventional total tidal disruptions will have spectra similar to common total disruptions in virtually every instance.  To the degree that their luminosity is tied to mass fallback rate, they may also have broader peaks and slower declines.

We have also shown that partial disruptions can alter the orbital evolution and induce mass evolution of the stars involved.  In particular, they drive evolution of orbital energy much faster than gravitational encounters with other stars.  This can, in turn, change the pace of the stars' angular momentum evolution.  The induced mass evolution results in eliminating high-mass stars from the population that is ultimately totally disrupted when their angular momentum evolves slowly.  Because partial disruptions temporarily distend the remaining star by  converting what had been warm mid-layers into outer envelopes \citep{Ryu3+2019}, an unconventional total disruption may take place only a single orbit after a severe partial disruption.

\section{ Summary}

 We have characterized the results of different TDEs depending on their pericenter distances; the relative rates of TDEs of different character can be determined from the star's angular momentum distribution. We summarize here  our main findings. 

\begin{itemize} 

\item ``Circularized" TDEs, events that follow the classical picture of rapid circularization and formation of a small accretion disk, are only a small fraction of TDEs for typical SMBHs ($M_{\rm BH}\lesssim 5 \times 10^6$) .  This is so both because these events can occur only for small stellar angular momentum and because the critical angular momentum is not much greater than the value within which direct capture occurs.  In fact, in most of these cases, the disk is still highly eccentric, although less so than the initial debris orbits.
The rate is suppressed further if the stellar angular momentum distribution has an empty loss-cone, in which the stars slowly diffuse in angular momentum phase space.
For heavier SMBHs, the fraction increases somewhat, but is significantly smaller than the one found by \cite{Guillochon.Ramirez-Ruiz2015} because of direct capture.

 \item ``Common" or ``slowed" TDEs don't follow the classical picture, but should comprise a clear majority of events.  
These take place when the
pericenter is greater than  $\simeq R_{\rm circ}$, a quantity almost independent of $M_*$ and $M_{\rm BH}$, but subject to uncertainty due to the poorly-known properties of eccentric-orbit accretion.
In this case, stream intersection takes place so far from the SMBH that the associated shock dissipates too little energy to bring the stream into a compact accretion disk surrounding the SMBH \citep{Shiokawa+2015,Guillochon.Ramirez-Ruiz2015,Dai+2015}.  Consequently, mass accretion onto the SMBH doesn't follow the mass fallback rate.  On the other hand, following the numerical simulations of \cite{Shiokawa+2015}, \cite{Piran+2015} suggested that the initial UV-optical luminosity can be supported by heating due to ``outer shocks", and this luminosity would approximately follow the mass fallback rate.

 \item As common or even more frequent than  the ``common" TDEs are ``partial" TDEs, in which the star loses only a fraction of its mass. The mass fallback rate in these cases is, of course, smaller and generally steeper than in full disruptions.  If the stellar angular momentum distribution evolves in the empty loss-cone regime, these will be the majority of all events and must precede all total disruptions.

\item Upon returning to the vicinity of the SMBH, a partial disruption remnant may be disrupted again. If the remnant did not relax to a main sequence-like configuration, the next disruption may produce a very peculiar event.  Both the rate and the fate of such events are uncertain and require further calculation of the rate of thermal relaxation of the stellar remnant and the evolution of its orbital energy by interaction with other stars surrounding the SMBH.

\end{itemize}

\section*{Acknowledgements}
{ We thank Nicholas Stone for helpful discussions.} This work was partially supported by NSF grant AST-1715032 and an advanced ERC grant TReX. 

\vskip 1cm

\appendix
\section{Potential Kerr Effects}

Black hole spin can potentially affect some of our estimates in two ways.

First, the apsidal precession depends upon spin.  Following \cite{Merrittbook2013}, \cite{Guillochon.Ramirez-Ruiz2015} point to two higher-order post-Newtonian contributions, which they designate as due to Lense-Thirring and quadrupole effects.  The ratio of the Lense-Thirring quadrupole contribution to the lowest-order post-Newtonian (Schwarzschild, or in their terminology, ``de Sitter") contribution is
\begin{align}
    \frac{\Delta \phi_{\rm aps}^{LT}}{\Delta\phi_{\rm aps}^S}& = -\frac{16}{3 \cdot 2^{3/2}} (a/M) \cos i (r_p/r_g)^{-1/2}\nonumber\\ &\simeq  -0.6 (a/M)\cos i (r_p/10r_g)^{-1/2}.
\end{align}
In other words, in the vicinity of our estimated $R_{\rm circ}$, its relative contribution is at most $0.6\times$ the Schwarzschild prediction, is generally smaller by the product of $(a/M) \cos i$, and, when averaged over an isotropic event population, is zero.  The same ratio for the quadrupole term is
\begin{align}
    \frac{\Delta \phi_{\rm aps}^{Q}}{\Delta\phi_{\rm aps}^S}& = (1/8)\left(1 -  5\cos^2 i\right)(a/M)^2 (r_p/r_g)^{-1}\nonumber \\
    &= 0.0125 \left(1 - 5\cos^2 i\right)(a/M)^2 (r_p/10 r_g)^{-1}.
\end{align}
Thus, for $r_p \gtrsim 10r_g$, this term is always $\lesssim 0.01\times$ the Schwarzschild contribution.

If the debris orbits are inclined with respect to the black hole spin, they are also subject to Lense-Thirring precession. If the stream is narrow enough, and $\Delta\omega$, the precession angle, is large enough, the stream may evade a close-in intersection \citep{Guillochon.Ramirez-Ruiz2015}. Because the initial debris orbits are so eccentric, the precession per orbit is well-approximated by the precession occurring near pericenter:
\begin{equation}
\Delta\omega\simeq  0.14 (a/M) (r_p/10r_g)^{-3/2}.
\end{equation}
 The stream is strongly heated every time it passes through the nozzle shock on the way to pericenter.  The resulting free expansion causes its width to grow substantially \citep{Guillochon.Ramirez-Ruiz2015}.  Over the time required to transit the pericenter region, the angle occluded by the stream grows to
\begin{equation}
    \frac{W}{r_p} = \frac{v_e}{c} (r_p/r_g)^{1/2} \simeq 6.3 \times 10^{-3}(r_p/10r_g)^{1/2}.
\end{equation}
Here $v_e$ is the stellar escape speed (equivalent to its characteristic internal sound speed); this is $\simeq 600$~km~s$^{-1}$ with relatively little variation across the main sequence.
Self-intersection can be avoided during the first pericenter passage if $\Delta \omega > W/r_p$, or
\begin{equation}\label{eq:LTcrossing}
    \frac{\Delta \omega }{W/r_p} \simeq  22 (a/M)(r_p/10r_g)^{-2} > 1.
\end{equation}
Thus, this condition is satisfied for streams with pericenters near $R_{\rm circ}$ if the black hole spins rapidly, but fails if $r_p \gtrsim  50(a/M)^{1/2}r_g$.

However, as shown by \cite{Dai+2015}, when $r_p \gtrsim 10r_g$, the point of intersection can be at a radius considerably larger than $r_p$.  In that case, the stream has additional time to expand before passing near another stream.  In fact, the simulation of \cite{Shiokawa+2015} shows continued expansion over much of the stream's orbit.  If it does continue, the ratio of nodal precession angle to the stream opening angle at the intersection point  falls $\propto (r_p/10r_g)^{-3/2} (r_{\rm int}/r_g)^{-1/2}$, where $r_{\rm int}$ is the distance from the black hole at intersection.  Another fit to the results of \cite{Dai+2015} indicates that $r_{\rm int} \propto r_p^{2.5}$, much like $a_{\rm orb}$ (see Sec.\ref{subsec:structure}), so that the overall scaling is roughly \begin{equation}
     \frac{\Delta \omega}{W/r_{\rm int}} \approx  4 (a/M) (r_p/10r_g)^{-11/4}.
\end{equation}
The coefficient is smaller than in the previous expression (Eqn.~\ref{eq:LTcrossing}) because $r_{\rm int} > r_p$ even for $r_p=10r_g$.  Thus, stream intersections due to Lense-Thirring precession rapidly become more dependent on the details of free expansion as $r_p$ increases beyond $\sim 10r_g$.

Similarly, continued free expansion beyond the vicinity of $r_p$ also means that when the stream returns to the pericenter region, even if it had been narrow enough to avoid intersection during the first passage, it is much wider than when it left.  During this second visit it is then likely to no longer meet the criterion of Equation~\ref{eq:LTcrossing}.

In sum, nodal precession may have relatively little effect on orbits with $r_p$ noticeably greater than $\gtrsim 10r_g$, while for pericenters $\lesssim 10r_g$, nodal precession may help streams evade intersection for at least one orbit. For the purpose of our argument, what matters is that it does not systematically increase the probability of a strong shock once intersection does occur.

%\bibliography{biblio,grantrefs,tde}

\end{document}